\documentclass[fleqn,10pt]{wlscirep}
\usepackage[utf8]{inputenc}
\usepackage[T1]{fontenc}
\title{Asymmetric participation of defenders and critics of vaccines to debates on French-speaking Twitter}

\author[1,+]{Floriana Gargiulo}
\author[2]{Florian Cafiero}
\author[1,3]{Paul Guille-Escuret}
\author[3]{Valérie Seror}
\author[1,3,*,+]{Jeremy Ward}
\affil[1]{CNRS, Université Paris Sorbonne, GEMASS, 75017, Paris, France}
\affil[2]{Université Sorbonne Paris-Cité, CNRS, LIED, 75013, Paris, France}
\affil[3]{Aix Marseille Université, IRD, AP-HM, SSA, VITROME, 13005, Marseille, France}

\affil[*]{jeremy.ward.socio@gmail.com}

\affil[+]{these authors contributed equally to this work}


\begin{abstract}
For more than a decade, doubt {about vaccines has become an increasingly important global issue}. Polarization of opinions on this matter, especially through social media, has been repeatedly observed, but details about the balance of forces are left unclear. In this paper, we analyse the flow of information on vaccines on the French-speaking realm of Twitter between 2016 and 2017. Two major asymmetries appear. Rather than opposing themselves on each vaccine-related controversy, pro- and anti-vaccine accounts focus on different vaccines and vaccine-related topics. Pro-vaccine accounts focus on hopes for new groundbreaking vaccines and on ongoing outbreaks of vaccine-preventable illnesses. Vaccine critics concentrate their posts on a limited number of “controversial” vaccines and adjuvants. Furthermore, vaccine-critical accounts display greater craft and energy, using a wider variety of sources, and a more coordinated set of hashtags. This double asymmetry can have serious consequences. Despite the presence of a large number of pro-vaccine accounts, some arguments raised by efficiently organized and very active vaccine-critical activists are left unanswered.
\end{abstract}
\begin{document}
\flushbottom
\maketitle

\thispagestyle{empty}
{ For more than a decade now, public doubt about vaccines has become an increasingly important global issue} \cite{dube2013vaccine, larson2016state}. This has recently led the World Health Organization to include “Vaccine Hesitancy” – i.e. negative attitudes towards vaccines that do not amount to a radical refusal of any form of vaccination – in its list of “ten threats to global health in 2019” \cite{larson2019reverse}.

The emergence of the Internet and its virtual social networks has played an important role in this global phenomenon \cite{betsch2012opportunities, marshall2018vaccine, pananos2017critical}. Since their inception, vaccination campaigns have generated some resistance from part of the public and the formation of groups of vaccine-critical activists \cite{durbach2004bodily, colgrove2006state, conis2015vaccine, blume2017immunization, poland2011age}. The invention of the Internet has provided these activists with new opportunities to reach a wider audience, beyond their traditional radical constituency. Studies {conducted at the beginning of the 2000s showed that vaccine critics were already on the Internet and that their arguments were very easily accessible via keywords queries in mainstream search engines} \cite{nasir2000reconnoitering, zimmerman2005vaccine}. They have since been very active on most prominent virtual platforms ranging from chatrooms, social media such as Facebook, Youtube or Instagram, and comments sections below mainstream media articles \cite{ward2016vaccine, schmidt2018polarization, atlani2015blood}. Analysts suggest that the features of online platforms – especially online social networks centred on “virality” - favour the spread of their arguments \cite{betsch2012opportunities, marshall2018vaccine, lewandowsky2017beyond}. This has led many experts to present the rise of vaccine hesitancy as the perfect example of how the Internet facilitates the spread of «fake news», « conspiracy theories » and a general shift towards a « post-truth society » \cite{marshall2018vaccine, lewandowsky2017beyond, betsch2017advocating, iyengar2019scientific, scheufele2019science, ferrara2016rise, broniatowski2018weaponized}.

Those who ought to defend vaccination - public health authorities, medical professionals, academics, internet giants – have long been accused of not doing enough to counter the spread of vaccine misinformation on the internet. But in the past years, these calls for action seem to have been heard: several online platforms have taken measures to decrease the virality of vaccine-critical contents \cite{larson2019reverse}, national authorities and researchers have developed and applied new online communication tools \cite{marshall2018vaccine, larson2018state}, and pro-vaccine social movements have emerged in several countries with citizens choosing to devote part of their free time to convince hesitant parents or to defend science more generally \cite{vanderslott2019exploring, ward2019antivaccins}. Recent studies suggest that, thanks to these mobilizations, the internet is not the realm of vaccine critics anymore \cite{schmidt2018polarization, ferrara2016rise, lutkenhaus2019mapping} and, in some regions, pro-vaccine messages may be gaining the upper hand \cite{menczer2019conversation}.

{There is little doubt} that more pro-vaccine presence on the internet is a positive development. In a context where a significant portion of Americans still believe that the MMR vaccine causes autism, it is crucial that vaccine critics’ public statements are not left unchallenged \cite{marshall2018vaccine}. But more presence does not necessarily mean that vaccine critics’ arguments are consistently debunked. {the algorithms governing properties built in digital platforms facilitate the constitution of filter bubbles and echo chambers which can lead to polarized debates and audiences} \cite{marshall2018vaccine, schmidt2018polarization, ferrara2016rise, broniatowski2018weaponized}.
{For instance, it is possible that both camps talk about different  vaccines or aspects of vaccination. Several specialists of vaccine-related controversies and vaccine hesitancy have drawn on qualitative case-studies to suggest that this is often the case in discussions of vaccines \cite{reich2018calling, colgrove2006state, conis2015vaccine}. But, to our knowledge, this aspect of public discussions of vaccines has not yet been the object of a broader research program.}\

{In this paper, we begin the work of testing whether there are such differences in thematic focus between vaccine defenders and critics using broader datasets. We analyse debates on vaccines that took place in French on Twitter between March the 28th 2016 and May the 5th 2017. We also assess the importance of these differences by studying the relative influence of the two camps.} 
{The French-speaking world, and France more specifically, is of particular interest to the study of doubts about vaccines, as France is one of the most vaccine hesitant countries in the world \cite{ward2019vaccine} and hesitancy is also particularly prevalent in Quebec, Switzerland and Belgium \cite{rowe2019does,dube2019overview}. This period is important because it precedes the election of Emmanuel Macron as president of France. One of the first policy announcements of his government was the extension of mandatory vaccination of children from 3 to 11 vaccines. This decision was meant to put an end to a difficult decade for vaccines in France. Since 2009, controversies over the safety of vaccines have continuously made the news, the French have become one of the most vaccine hesitant populations in the world and France has seen several measles epidemics \cite{ward2019vaccine}. As in many other countries or states such as Italy, Germany, Australia or California, French public health authorities have resorted to constraint in the hope of raising vaccination coverage. Because, in some cases, resorting to constraint can backfire by stimulating the constitution of organised anti-vaccine movements or a more general lack of trust in authorities \cite{durbach2004bodily, blume2017immunization, ward2017france, leask2017imposing, bedford2014pro}, it is crucial to better understand the context leading to such decisions. This helps to identify the limits of existing communication strategies and determine whether coercion is inevitable – while keeping in mind the fact that communication and hesitancy are just one part of the puzzle of low vaccination coverage} \cite{omer2019mandate}.\

{We focus on the social platform Twitter because it is the data source best suited to our research question. Each social media has its specificities stemming from its built-in properties and the social profile of its users. While twitter is known to be used relatively more by members of the middle and upper-classes, it is also known to be heavily invested by activists of all types and institutions (at least in France \cite{boyadjian2017conditions}. Its features (user activity, number of followers, of retweets, etc.) also allow distinguishing between activists and less engaged members of the public, contrary to a relatively more widely used social media such as Facebook.Twitter is therefore well suited to research focusing on the interaction between activists and the public.}\

{To test the hypothesis of a difference in thematic focus between vaccine defenders and critics, we analyse the structure of the retweet network for each vaccine-related topic and for all topics taken as a whole. We find that most childhood vaccines arouse a similar degree of interest in critics and defenders but that their respective productions tend to focus on different keywords. We also find that some issues are almost completely abandoned by pro-vaccine actors even though they can be at the core of contemporary vaccine hesitancy (adjuvants and additives for instance). To assess the importance of these differences, we analyse the relative centrality of critics and defenders in each retweet network as well as their relative reach using a variety of measures. We show that, ceteris paribus, pro users are retweeted by a larger number of accounts but that critics tend to use the specific functionalities of Twitter better even though it doesn’t go so far as to give them a wider reach than vaccine defenders.}

\section*{Data}
{Using a combination of the streaming and search Twitter API, we collected all tweets pertaining to vaccination published in French between March 28\textsuperscript{th}, 2016 and May 23\textsuperscript{rd}, 2017 (258.166 tweets posted by 107.923 unique users, see Supplementary material for the full list of keywords used in the data collection). In this paper, we focus on the 58.559 tweets from 31.088 unique users dealing with specific vaccines (existing or hotly awaited), i.e. containing the name of a commercial vaccine or the name of a vaccine-preventable disease and variants of the term "vaccine". We therefore excluded tweets about vaccination in general. We listed 37 vaccines or substances contained in vaccines based on the commercial vaccines available in France, Belgium, Canada and Switzerland, and on our knowledge of existing controversies in these countries (see Table 1 for full description of the 111 keywords used). These 37 topics fall into 5 categories: seasonal flu, mandatory and recommended for the general population in developed countries, adjuvants and additives, hotly awaited, and other vaccines (for rarer diseases, developing countries or specific subgroups). In addition to being necessary to answer our research questions, focusing on this more precise set of keywords had the effect of limiting the risk of integrating irrelevant tweets {More details on the data collection and cleaning are reported in the supplementary material file.} The keywords are presented in Figure}\ref{fig:kwPrev}. \\

\begin{table}
\begin{tabular}{|l|l|p{5cm}|}
\hline
                  CATEGORY &                     TOPIC &                                           KEYWORDS \\
\hline
                  SEASONAL &                       flu &                                             grippe \\\cline{2-3}
                   &           vaccine\_for\_flu &                             vaxig; fluar; influvac \\
\hline
 MANDATORY\_AND\_RECOMMENDED &      human\_papillomavirus &   du col; uterus; papllo; hpv; papilo; vph; col de \\\cline{2-3}
  &                   rubella &                                           roubeole \\\cline{2-3}
  &              pneumococcal &                                             pneumo \\\cline{2-3}
  &         vaccine\_papilloma &                             cervar; gardas; gardaz \\\cline{2-3}
 &                   tetanus &                                              tetan \\\cline{2-3}
 &               hepatitis b &   hepb; hepattite b; hepatitte b; hep b; hepatite b \\\cline{2-3}
 &                   measles &                                           rougeole \\\cline{2-3}
 &    haemophilous influenza &                               type b; haemo; typeb \\\cline{2-3}
 &  vaccine\_mening\_pneumonia &             prevenar; neisvac; pneumova; menjugate \\\cline{2-3}
 &        vaccine\_hexavalent &    revax; dtp; hexa; revaxis; repeva; enger; hexyon; tetra; boostri; genhevac; hbv; vaxelis; infar; penta \\\cline{2-3}
 &                 diphteria &                                              dipht \\\cline{2-3}
 &                     mumps &                                          oreillons \\\cline{2-3}
 &               mmr\_vaccine &                                  ror; priorix; mmr \\\cline{2-3}
 &                     polio &                                              polio \\\cline{2-3}
 &            whooping\_cough &                                             coquel \\\cline{2-3}
 &              tuberculosis &                                         tuber; bcg \\\cline{2-3}
 &                meningitis &                             mening; menc; menningi \\
 \hline
   ADJUVANTS\_AND\_ADDITIVES &                 adjuvants &                    alu; conservateur; adjuv; squal \\\cline{2-3}
  &                 additives &          thime; adi; conserv; thiome; mercur; addi \\
  \hline
                     OTHER &                 rotavirus &                                               rota \\\cline{2-3}
 &               chicken pox &                                          varicelle \\\cline{2-3}
&            vaccines\_other &    tyavax; ixiaro; twinrix; variva; vaqta; menveo; bex; havrix; stamaril; varilri; rotarix; spirol; nimenr; tico; typheri; avaxim; typhim vi; zostava \\\cline{2-3}
&              yellow fever &                                       fievre jaune \\\cline{2-3}
 &                      lyme &                                               lyme \\\cline{2-3}
 &                  shingles &                                               zona \\\cline{2-3}
 &         hepatitis\_a\_and\_c &    hepa; hepc ; hepatitec; hepatitea; hep a; hep c; hepatite c; hepatite a \\\cline{2-3}
&             leptospirosis &                                              lepto \\\cline{2-3}
 &                   cholera &                                            cholera \\\cline{2-3}
 &              thyfus fever &                                           thyphoid \\\cline{2-3}
 &              encephalitis &                                        encephalite \\
 \hline
                   AWAITED &                     ebola &                                              ebola \\\cline{2-3}
 &                      aids &                               sida; aids; vih; hiv \\\cline{2-3}
 &                    dengue &                                             dengue \\\cline{2-3}
&                   malaria &                                 paludisme; malaria \\\cline{2-3}
 &                      zika &                                              zika  \\
\hline

\end{tabular}
\caption{\label{kwClass}Keywords, topics and categories}
\end{table}

\section*{Results}
\begin{figure}[h!]
\centering
\includegraphics[width=0.6\textwidth]{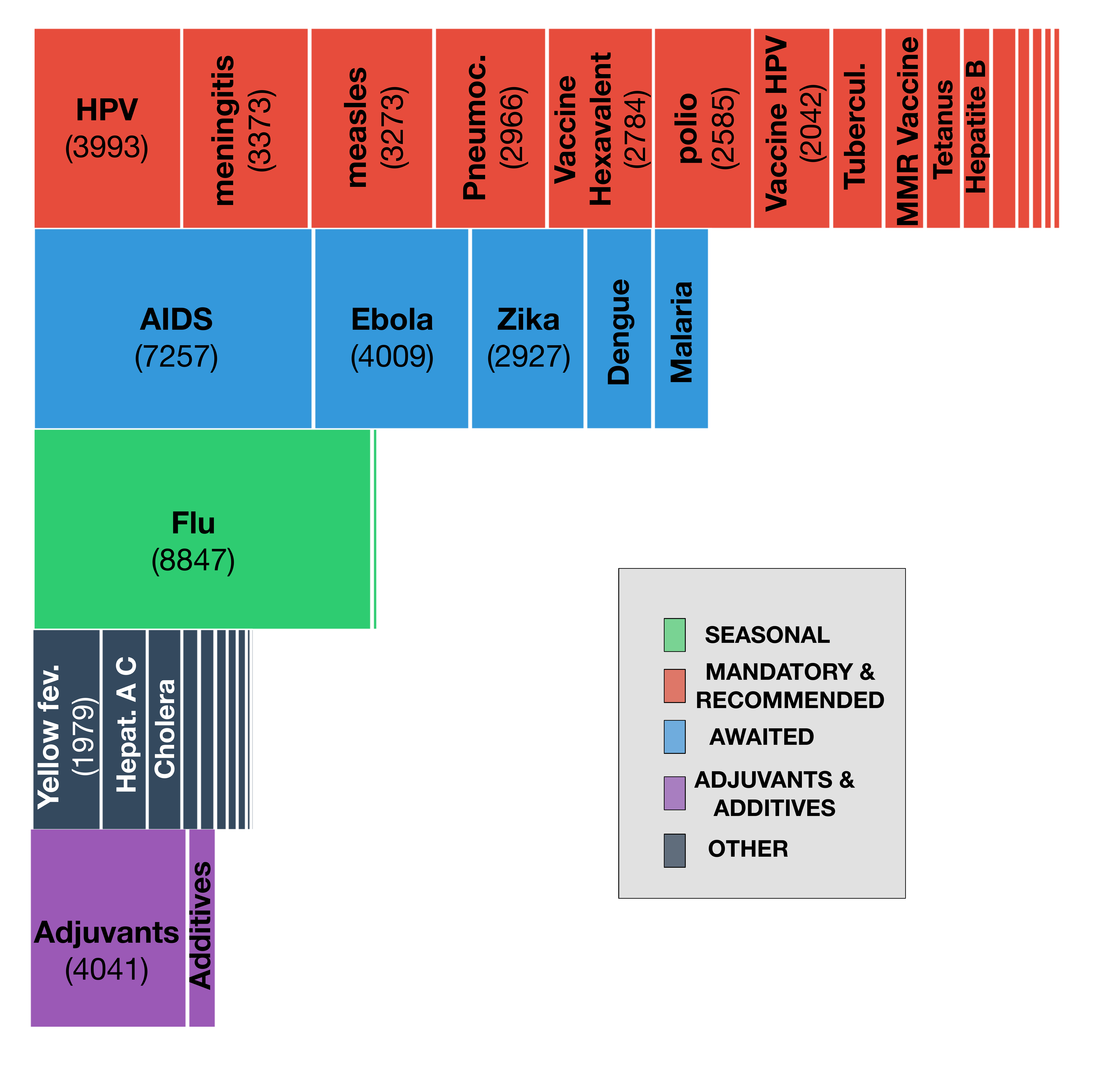}
\caption{Volume of tweets by topic}
\label{fig:kwPrev}
\end{figure}

{Between March 2016 and May 2017, the most discussed vaccine on the French Twitter feed was the seasonal flu, as we can observe in Fig.\ref{fig:kwPrev}. A large attention was also dedicated to 'awaited' vaccines, like the vaccines against AIDS and  Ebola. Among our five groups of topics, "Recommended and Mandatory vaccines" had the most content. Adjuvants and additives were less debated than our other groups of topics even though they have been at the center of heated debate in France for the past ten years.\\
To test our hypothesis, we will first present our analysis of our whole sample of tweets and then focus on each vaccine-related topic taken separately. Finally, we will focus on the respective reaches and patterns of activity of vaccine critics and defenders.}

\subsection*{{Vaccine critics and defenders tend to focus on different vaccines}}

\subsubsection*{{The circulation of information on all vaccines}}
\begin{figure}[h!]
\centering
\includegraphics[width=\textwidth]{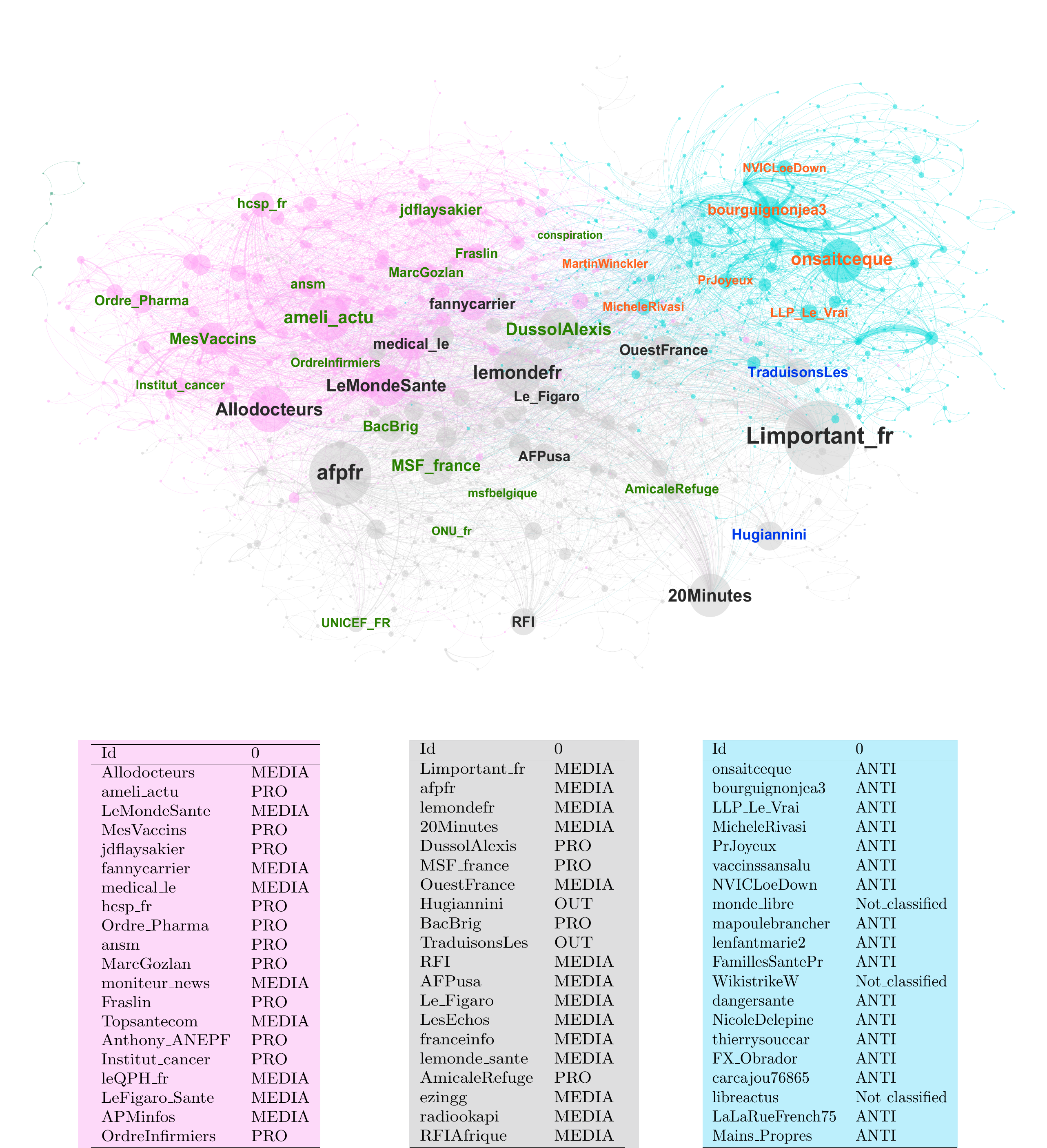}
\caption{Retweet network. The color of the nodes corresponds to the community structure retrieved trough the Louvain algorithm. The color of the nodes' labels corresponds to the classification of the users: "ANTI" (orange), "PRO" (green), "MEDIA" (grey). The size of the node is related to the node's in-degree. The table contains the users with highest in-degree with their classification, for each community.}
\label{rtAll}
\end{figure}

We partitioned the retweet network using the Louvain algorithm, {finding that a structure with 3 major communities is the partition maximizing modularity. As we can see from Fig.\ref{rtAll}, the retrieved communities show some similarity with the users' categories}. One community (light blue) contains the most important "ANTI" users ; another one (rose) several important "PRO" users. The third one (grey) contains users  mostly related to media accounts and to the NGO world. Notice however that while the community classification almost perfectly maps the "ANTI" users, several "PRO" accounts, related to institutions or health related media, are classified in the last two groups. 

\subsubsection*{{The role of defenders and critics in the flow of information on each vaccine}}
\begin{figure}[h!]
\centering
\includegraphics[width=\textwidth]{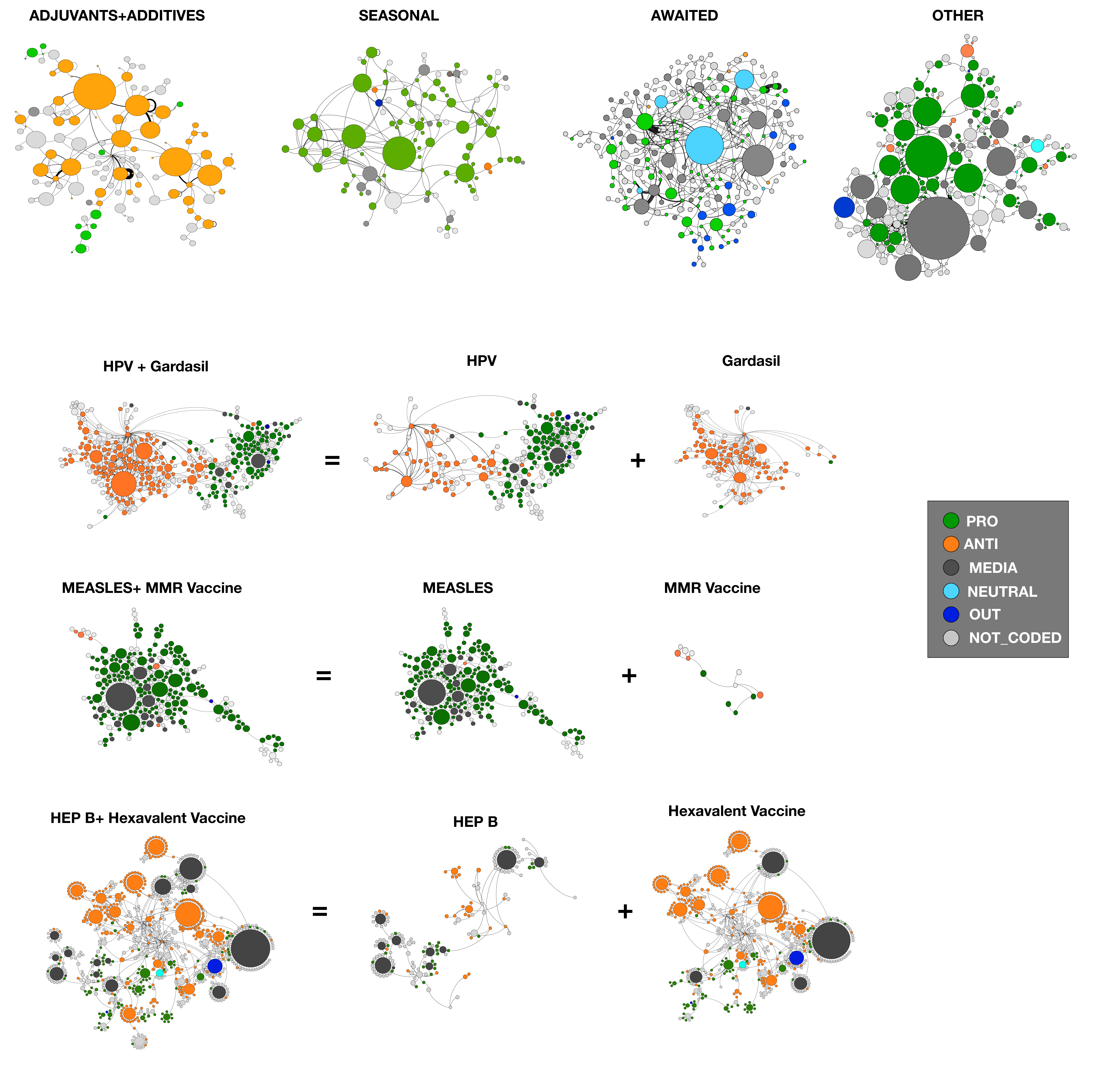}
\caption{Layers of the retweet network for four categories (ADJUVANTS, SEASONAL FLU, AWAITED and OTHER vaccines) and 3 controversial subjects (HPV, Measles, Hepatitis B). The color of the nodes corresponds to the classification of the users: "ANTI" (orange), "PRO" (green), "MEDIA" (dark grey), "NEUTRAL" (cyan), "un-classifiables" (dark blue) and not coded (light grey). }
\label{rtTopics}
\end{figure}

To get a more detailed view of the retweet network, we constructed a multiplex network, associating a layer to each topic. In some cases, we further decomposed the layer structure at the topic level. Some example of the layers are reported in Fig.\ref{rtTopics}.

A vast majority of topics is mostly broached by one side only. {The flow of information} on adjuvants and additives is for instance dominated by vaccine critics, while the flows on flu and measles are dominated by pro-vaccine users.  In rare cases, critics and defenders seem to discuss exactly the same topic, like the human papillomavirus (HPV). This opposition is however not exactly symmetrical: HPV as a vaccine-preventable disease is mainly discussed by pro-vaccine users ; but the actual HPV vaccines (GardasilTM and CervarixTM) are a topic almost exclusively broached by vaccine critical users.

\begin{figure}[h]
\centering
\includegraphics[width=\textwidth]{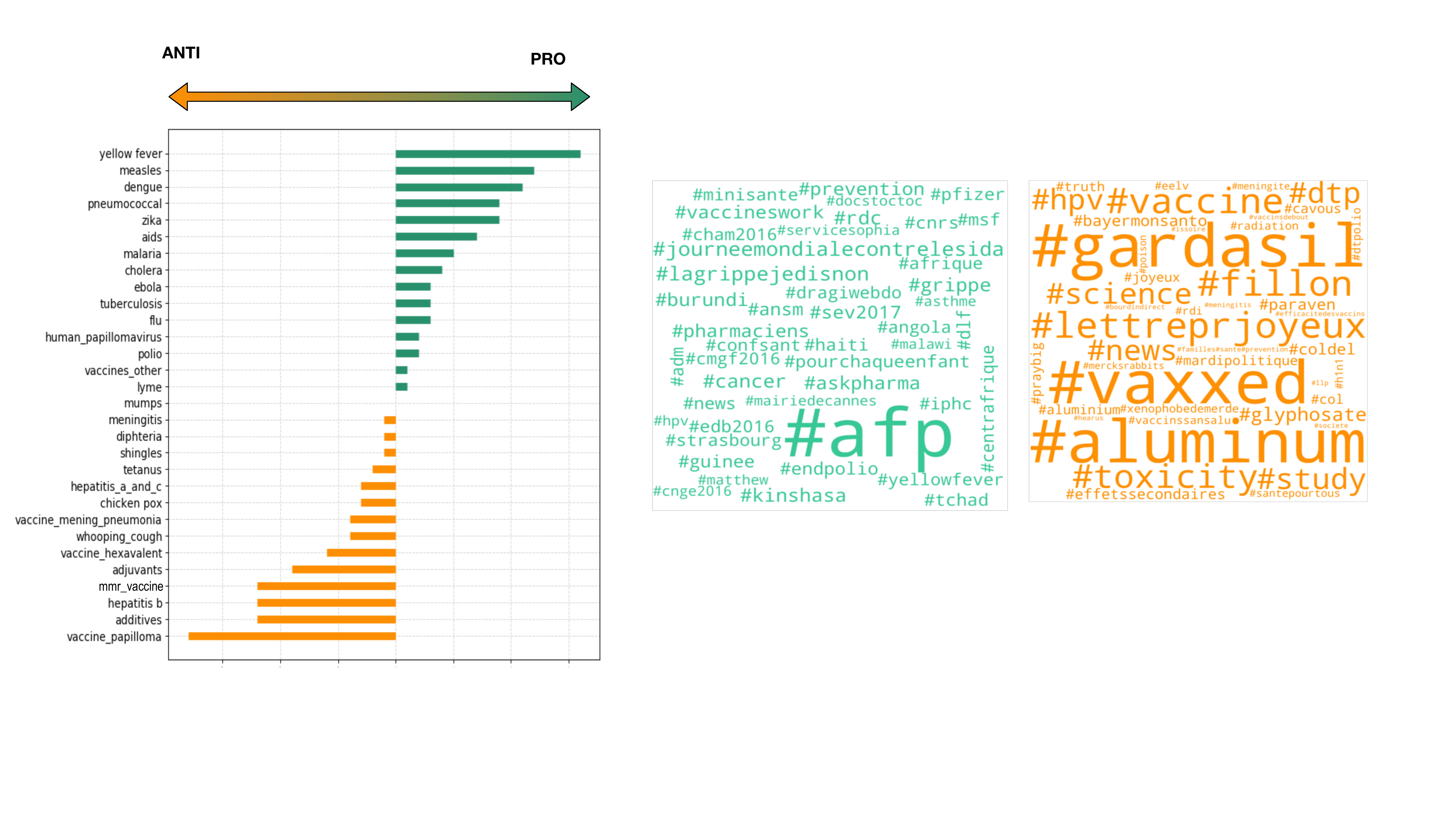}
\caption{Most representative topics for PRO and ANTI classes. Word clouds for PRO and ANTI classes. }
\label{colors}
\end{figure}

A final step to define the topic preference in the "PRO/ANTI" classes is to go back from the topic retweet graphs. For each topic, $\alpha$, we calculated the number of tweets citing the topic for each user in the activist list, $A_{Ext}$ and $P_{ext}$. Based on this value, we reconstructed the ranking of the topics in each class ($r_{PRO}(\alpha), r_{ANTI}(\alpha)$), and its relative ranking $R(\alpha)=r_{PRO}(\alpha)-r_{ANTI}(\alpha)$: a strongly positive (resp. negative) relative rank indicates a mostly "PRO" (resp. "ANTI") topic, a small value of this measure indicates a neutral topic, broached almost equally by the two camps. The relative ranking of the most cited topics is reported in the lower plot of Fig.\ref{colors}. The war horses of the "PRO" users are yellow fever and other tropical diseases, the diseases associated to mandatory vaccines and the seasonal flu. {The "ANTI"-vaccine class is strongly focused on the limited number of vaccines and substances that have become very controversial, at least in France such as adjuvants and the hepatitis B vaccine (additional measures of polarization for each topic are reported in the supplementary information file)}.

\subsection*{{Vaccine defenders reach a wider audience but vaccine critics use Twitter more effectively}}
{We will now analyze the potential of  pro-vaccine and vaccine-critical accounts to reach a large public and to influence opinion.}

\subsubsection*{{The reach of vaccine defenders and critics}}
\begin{figure}[h!]
\centering
\includegraphics[width=0.9\textwidth]{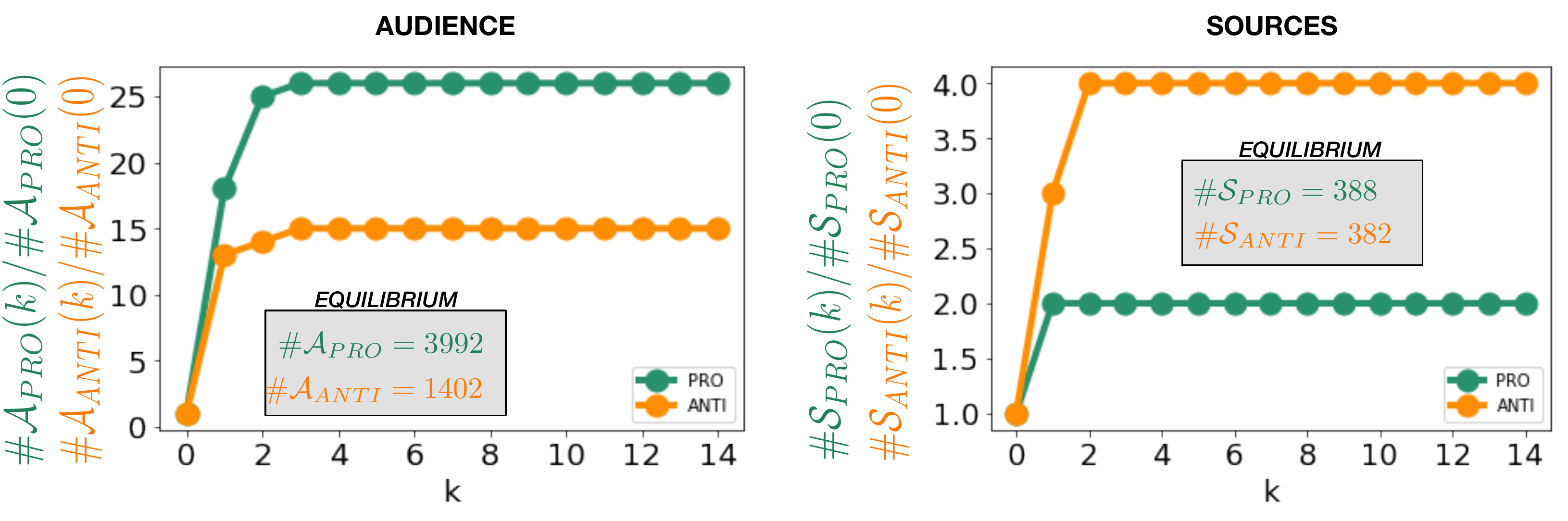}
\caption{Left plot: size of the $k$-shell audience for the "PRO" (in green) and of the "ANTI" (in orange) set, normalized by its initial size. Right plot: size of the  $k$-shell sources for the "PRO" (in green) and of the "ANTI" (in orange) set, normalized by its initial size.}
\label{audience}
\end{figure}

In figure \ref{audience}, we represent the size of the $k$-shell audience and of the $k$-shell sources for the two groups of activists. Since the size of the original groups was not equal, we divided the sizes of the $k$-shell audience (sources) by its initial value in order to compare the growth mechanisms. 
{The $k$-shell audience represents all the users that exclusively shared the information from one of the two groups. Similarly the $k$-shell sources are all the users who were cited by only one of the two groups. Once normalized by the initial size, these normalized measures represent respectively the average capacity of each user in a group to be retweeted (for the audience) and its average retweeting activity  (for the sources).} We can observe that, for the "PRO" community the audience size stabilizes at a higher relative value. \textit{Ceteris paribus}, the "PRO" users' posts are retweeted by a larger number of accounts

{Iterating the shell definition procedure until all users are included, we observe that the audience of the “PRO” and “ANTI” cover together 32\% of the retweet network nodes. However,  pro--vaccine activists are capable of reaching 24\% of the users, while the —vaccine critics only reach 8\%. We can observe that in the remaining pool of users (68\% who do not exclusively retweet one or the other), only 0.6\% retweet both “pro” and “anti” users. The largest part of these users therefore only retweet contents produced by news media accounts}.\

\begin{figure}[h!]
\centering
\includegraphics[width=0.9\textwidth]{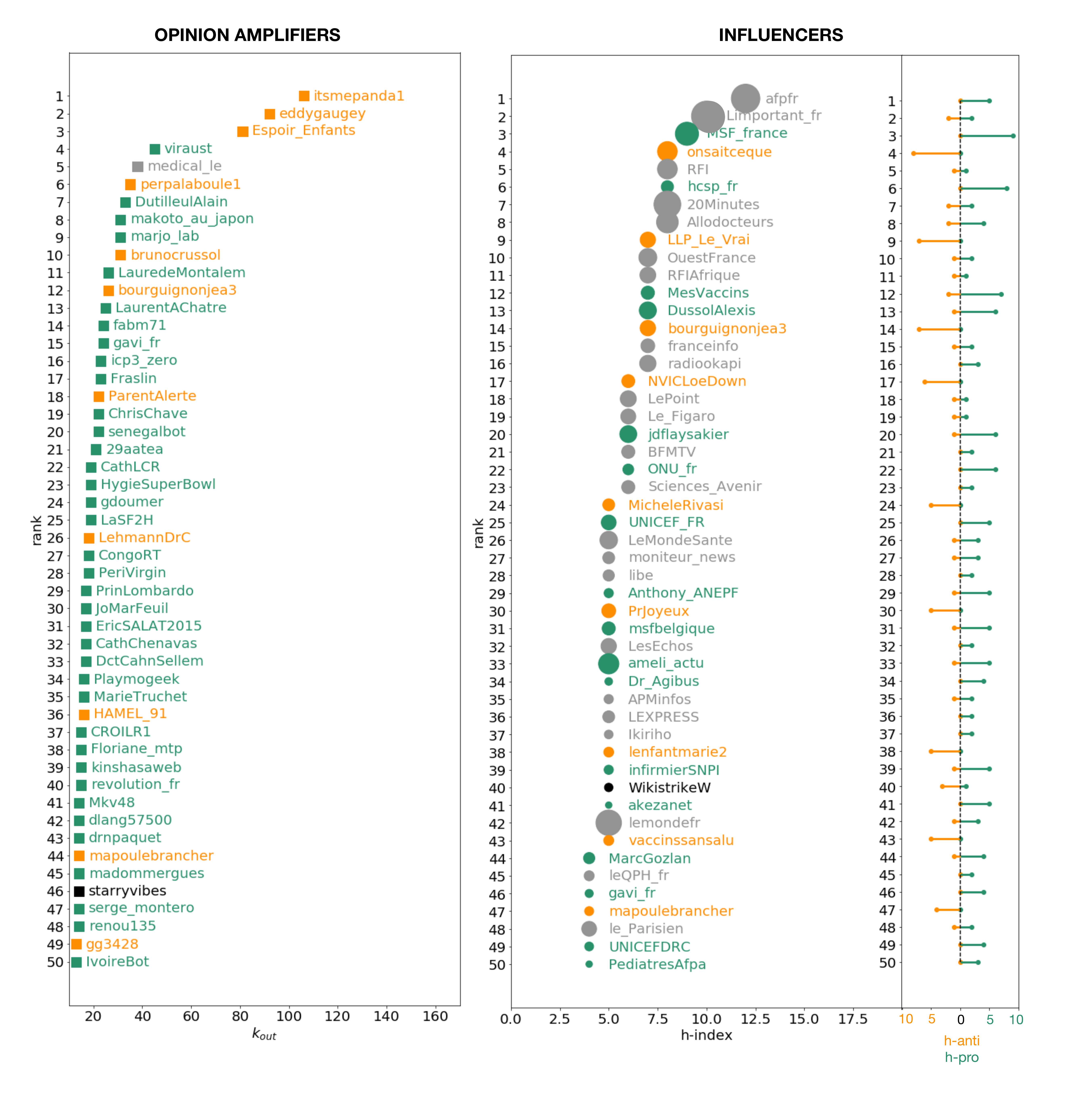}
\caption{{Left plot: top 50 users with the highest out—degree (amplifiers). "PRO" in green, "ANTI" in orange and "MEDIA" in grey. Right plot: top 50 users with the highest h—index (influencers). The size of the point corresponds to the number of retweeted posts. On the right we report for each user in the list of influencers, its associated h-score for the PRO and the ANTI communities.}}
\label{influential}
\end{figure}

{In order to better understand this result, and to assess the importance of each type of actor in the spread of information, we used some centrality indicators. First we defined the users posting a large number of retweets (users with a large out—degree) as \emph{opinion amplifiers}}  $k_{out}$. {As we can see in the left plot of Fig.~\ref{influential}, a small number of vaccine critics present an unusually high retweeting activity and can be considered as super-amplifiers. But, descending at a lower activity level, the strongest amplifiers are vaccine defenders}. {As explained in the methods section, we assessed the level of influence of a user, using her h--index. The biggest influencers in the vaccine debate are clearly the newsmedia as we can observe in the right plot of Fig.~\ref{influential}. However, while most influential media are only authoritative for the PRO,  we observe that  a minority are equally retweeted by defenders and critics alike}.

\subsubsection*{{Critics are more active and use hashtags in a potentially more efficient way}}
{Vaccine critics have a more limited reach than vaccine defenders, but it is important to note that the difference between the two (24\% vs 8\%) is not that substantial given the large scientific consensus around vaccination and the resources at the disposal of some pro-vaccine actors (Ministries of Health, public agencies, scientific societies, pharmaceutical companies…). To understand this result, let us go back to our measures of the activity of these users. “Anti” users have on average a significantly higher tweeting activity, and make reference to a greater number of sources. "Anti" users tend to make the effort of trying to spread the messages they agree with, but are however less influential compared to the "Pros"}.

The most frequently used hashtags by pro-vaccine and vaccine-critical accounts presented in the right plots of  Fig.\ref{colors} suggest another difference in these two camps’ practices. {Defenders tend to disperse their use of hashtags over a greater number of them, to refer to themes that are too general to make their tweet stand out in the mass of contents posted on the subject (such as \#cancer, \#pharmaciens, \#Angola), to refer to institutions and media (for instance: \#afp, \#CNRS) and to focus on hashtags produced by institutions to mobilise around an official campaign (\#lagrippejedisnon,\#pourchaqueenfant, for example). The latter use of hashtag is efficient to mobilize a community of sympathiser but does not correspond to the keywords parents are likely to put in the search bar when they look for discussion on the actual vaccines they must make a decision on (see methods for a discussion of hashtag use). On the contrary, vaccine critics use both hashtags designed to mobilize their sympathizers (\#vaxxed, \#lettreprjoyeux…) and - more importantly - hashtags relating directly to specific vaccines or vaccine-specific substances (\#aluminium, \#hpv, \#gardasil) making it likely that their tweets will be found when someone searches information on a given vaccine}.

\section*{Discussion}
{In this paper, we partially corroborated the hypothesis that vaccine critics and defenders tend to focus on different vaccines. We found that most childhood vaccines attracted a similar amount of attention by critics and defenders but that their respective productions tended to focus on different keywords. We also found that some issues are almost completely abandoned by vaccine defenders even though they can be at the core of contemporary vaccine hesitancy (adjuvants and additives for instance). We also showed that pro users are retweeted by a larger number of accounts but that critics tend to use the specific functionalities of Twitter better even though it doesn’t go so far as to give them a wider reach than vaccine defenders}.

Our results contribute to current reflections on processes of polarization in the age of digital social media. The specific way social media are designed favours this process of polarization. By
suggesting new contents based on the users’ previous behaviour, they tend to create echo chambers
where the information circulating is culturally homogeneous thereby favouring partisan bias in the rare instances where users are put in contact with dissonant contents. This process is reinforced by producers of fake news who exploit social media’s orientation towards virality to make money from the
circulation of radically partisan contents \cite{lazer2018science}. This polarization not only affects perceptions of politicians’ actions. Scientific subjects are also caught up in these processes \cite{iyengar2019scientific}. Analysts have suggested that the current rise of vaccine hesitancy is at least partly due to a process of polarization \cite{schmidt2018polarization, iyengar2019scientific, broniatowski2018weaponized, larson2018biggest}. Few empirical studies have tested the level of polarization of discussions on social media. Schmidt et al. studied Facebook users who posted at least 10 posts on vaccines between 1st January 2010 and 31st May 2017 \cite{schmidt2018polarization}. They found that a majority of pro and anti-vaccine users only consumes and produces information in favour or against vaccines, not both, indicating a high degree of polarization. Menczner and Hui also found a very high degree of segregation between pro and anti-vaccine Twitter users\cite{menczer2019conversation}. {However, Lutkenhaus et al. did not find evidence of polarization in discussions of vaccination that took place during the second half of the year 2017 between dutch-speaking Twitter users \cite{lutkenhaus2019mapping}. The anti-vaccine community was largely connected to several pro-vaccine communities and pro and anti-vaccine users interacted regularly. Nevertheless, they also found that most of these interactions were conflictual (insults, mockery, criticism…). Our results prolong those of Lutkenhaus et al. and suggest that contents produced by Europeans tend to be less polarized than that produced in North America}.  
We found that vaccine critics and defenders composed two fairly cohesive communities on Twitter. But we also found that these two communities were significantly connected, at least indirectly, via their tendency to retweet the same mainstream newsmedia. The role of traditional newsmedia in polarization on social media is often overlooked even though they play a central role in political polarization \cite{lewandowsky2017beyond}. 

According to Benkler, Faris and Robert, the growing divide between right-wing and left-wing Americans is largely due to choices made by a number of media outlets – in connection with evolutions within the two main political parties- to use partisanship as a market strategy \cite{benkler2018network}. This logic presides over the constitution of digital echo chambers which are therefore only one of the many mechanisms through which these transformations affect the American public rather than the cause of polarization. France’s political landscape is much more multipolar and the French media landscape has not followed the same transformations. A recent study found that there remained a strong core set of agenda-setting elite media who adhere to a philosophy of journalistic objectivity and act as gatekeepers against fake news and radical views \cite{montaigne2019media}. The fact that both defenders and critics can turn to these mainstream media can be explained by the diversity of contents they have produced on the subject of vaccination in recent years. French journalists covering health can be divided when it comes to the legitimacy of concerns regarding the safety of vaccines on some specific issues (aluminium-based adjuvants, the safety of the HPV vaccine). But this division is often within each media’s newsroom rather than between media outlets \cite{ward2019journalists}. In addition to this, the outcomes of high-profile vaccine-related lawsuits, results of surveys showing the high levels of vaccine hesitancy in France, government decisions or public health officials’ statements regarding mandatory vaccinations regularly feature in all media. Consequently, this diversity of contents means that defenders of vaccines can retweet a piece published in Le Monde debunking common “antivaccine myths” while vaccine critics can retweet an interview of the head of a collective of “victims” of aluminium-based adjuvants performed a month earlier by another of Le Monde’s journalists. Our results therefore highlight the need to integrate knowledge of the diversity in newsmedia coverage of vaccination to interpret the structure of discussions on social media.

Our main contribution is our finding that defenders and critics of vaccines on Twitter focus on different topics, and especially on different vaccines. Vaccine critics mainly focus on the alleged dangers of specific vaccines. In the French-speaking world, these are the vaccines against HPV or hepatitis B, or adjuvants such as aluminium. But the list would likely be different in other cultural areas and countries as vaccine-related controversies tend to be grounded in local contexts \cite{dube2015vaccine}. Pro-vaccine accounts mostly focus on the dangers of a low vaccination coverage and on hopes raised by future vaccines. This asymmetry raises much concerns. Pro-vaccine accounts are numerous, and seem to attract a wider audience than vaccine critical accounts, which could be reassuring. Yet, vaccine-critics are very active and well coordinated   and some of their arguments are most of the time left unanswered by their opponents.

Our findings have implications for public health policy. Researchers working on vaccine hesitancy have argued that it is crucial to challenge vaccine critics’ arguments on social media and not let the Internet be the realm of antivaccinationism \cite{betsch2012opportunities, leask2017target}. How this should be done is currently the object of a heated debate. Some have warned against adversarial approaches and public stigmatization of vaccine critics which might make them appear as victims of persecution, suggest vaccination is a scientifically contested topic and increase polarization of attitudes \cite{marshall2018vaccine, bedford2014pro, leask2011target, silverman2017shaming}. Recent studies also suggest that debunking strategies can have counter-productive effects \cite{marshall2018vaccine, betsch2012opportunities, nyhan2015does}. In our study, we found that vaccine defenders talk less than critics about the more controversial vaccines or aspects of vaccination. Some of critics’ main arguments remain consistently unaddressed. It could mean that the strategy chosen by medical experts and other pro-vaccine actors consists in emphasizing the importance of the principle of vaccination for public health. This strategy has the advantage of not directly mentioning the objects of concerns which has been found to decrease vaccination intentions and to emphasize the importance of herd immunity which tends to alleviate doubts \cite{marshall2018vaccine, betsch2012opportunities, betsch2017benefits}. Nevertheless, in a context where an increasing number of vaccine critics present themselves as “not antivaxxers” and manage to convince both the public and journalists that they are different from traditional antivaccinationists \cite{reich2018calling, ward2019journalists}, it is doubtful that this approach will prompt a dismissal of their claims by the public. We believe that, even though more research is necessary to discover the best ways to debunk unfounded claims, defenders of vaccines should not wait for the discovery of a magic bullet before addressing these claims on social media – provided they follow simple ethical rules such as treating vaccine critics and hesitant parents with respect \cite{bedford2014pro, leask2011target, omer2019mandate }. 

That said, one of our results raises a new type of dilemma for vaccine communication. We found a strong presence of defenders in contents relative to vaccination against papillomaviruses. However, we also found that much of these contents centred on the market name of these vaccines (GardasilTM and,
more marginally, CervarixTM) and that vaccine critics completely dominated these contents. Should public doctors, experts and public authorities publicly defend a commercial product? In most developed countries, including France, public authorities are seen as being too close to pharmaceutical companies which contributes to a lack of trust in vaccines \cite{dube2013vaccine, larson2019reverse, bocquier2018social, peretti2019think, hobson2007trusting}. On the one hand, defending a specific commercial vaccine can reinforce the impression that financial interests bear on vaccination policies and market authorizations. But on the other hand, it is necessary to give reassurances that market
authorization processes are effective in assessing and monitoring the safety of vaccines.

{Finally, our study focused on an important period for debates over vaccines in France: the year preceding the announcement that the number of mandatory vaccines would be extended from 3 to 11 in June 2017. A lot has happened since then: the new mandate framework was put in place in January 2018, the French government launched an important communication campaign targeting the public and healthcare professionals and Media coverage of vaccine safety seems to have abated. Further investigation should be focused on whether and how these changes have affected the structure of the flow of information on Twitter described in this paper. Because, in some cases, resorting to constraint can backfire by stimulating the constitution of organised anti-vaccine movements or a more general lack of trust in authorities \cite{durbach2004bodily, blume2017immunization, ward2017france, leask2017imposing, bedford2014pro, omer2019mandate}, it is crucial to better understand the effect of these measures on public discussions and attitudes to vaccination}.

\section*{Limitations}
 
The main limitation of our analysis relates to the generalisability of our results. We focused on tweets written in French. Our results likely reflect the specificity of vaccine debates and vaccine hesitancy in France. This can be seen in the volume of discussions on aluminium-based adjuvants. The use of aluminium in vaccines has been at the core of most debates around vaccination in France since 2010 while it has not emerged as an object of major concern outside the French-speaking world \cite{ward2019vaccine}. Conversely, the dominance of vaccine defenders in discussions around the MMR vaccine could reflect the fact that this vaccine has not been the object of strong critical mobilisations in France, contrary to countries such as the United States of America and Great Britain \cite{peretti2019think, conis2015vaccine}. The idiosyncratic nature of vaccine hesitancy and of activists’ mobilisations on the subject of vaccination is likely to affect two parameters: a) which vaccines will attract most debate, and b) the overall balance of power between positive and negative discourses.	{Another limitation comes from our focus on Twitter. Because each social mediahas its specificities (practices and publics), it is possible that the types of contents and the flow of information differs radically on Facebook or Instagram for instance. However, we judge this to be unlikely as our results are coherent with the data available on vaccine hesitancy and vaccine-related controversies in France \cite{ward2019vaccine} and with data pertaining to discussions of vaccines on Twitter in other European countries \cite{lutkenhaus2019mapping}. Further research comparing the structure of the flow of information on vaccines on each social media and in different countries would contribute to current reflexions on the best ways to curtail the spread of misinformation on the Internet}.
\section*{Methods}
\subsection*{Retweet networks}
We mapped the information flows between users by following their retweeting activity: namely we constructed a directed graph where the nodes, $V_{RT}=[u_1...u_n]$, are the users and a directed edge is created between two nodes ($u_i,u_j$) if user $i$ retweets user $j$. The retweet graph, represents the circulation of information among users.
 A retweet most often means that a user endorses the idea expressed by the user she retweets. In this sense the retweet network can also be interpreted as an opinion similarity space. 
 We chose not to study "mentions". It has indeed been shown that, in the case of very polarized debates, like the US elections, polarization in terms of community structure is not observed in the mention graph \cite{conover2011political}. This comes from the fact that, in such contexts, mentions are often used to cite one's opponent.  
The giant component of the retweet network consists of 16.302 nodes, connected by 20.648 weighted directed edges - the number of retweets between two users defining a weighting. The full graph, including isolated nodes, is composed of 20.121 nodes and 23.348 edges.

{Using the topical tags listed in Table \ref{kwClass} we associated to each link of the retweet network the list of its associated topics (extracted from the text associated to the retweet):  In order to analyze how the information on each topic circulates, we analyzed the multi-layer structure of the retweet network: namely we decomposed the full graph in layers only containing links that are relative to a certain topic. It is a well documented fact that the aggregation of multiplex structures imply a significant loss of information that can strongly bias the observations \cite{menichetti2014weighted,battiston2014structural}. We investigated, in particular, how the nodes participate to the different layers and which roles they have in the overall network and in the multiplex structure}.

\subsection*{Identifying vaccine critics and defenders}
{We manually annotated 360 of the most active and/or prominent users on vaccination, noting their position toward vaccines (“pro”/”anti”/”neutral”) and whether they were accounts by newsmedia who covered the subject without expressing a personal point of view (coded as “media)}. 
The manual annotation procedure allowed us to identify a list of 92 vaccine critics ($A_{ini}$), 146 vaccine defenders ($P_{ini}$), 86 media and 36 neutral users.
From the initial sets of “Pro” and “Anti” users, we built the \emph{audience} and the \emph{sources} for the two groups, considering respectively the re-tweets pointing to each of the two groups and originating from them in the following way. \\

We first defined the 0-shell audience, for the two groups, as the initial sets:
$\mathcal{A}_{ANTI}^{(0)}=A_{ini},\mathcal{A}_{PRO}^{(0)}=P_{ini}$. At each iteration we calculated the $k$-shell audience adding to the $k-1$-shell audience the exclusive incoming neighbourhood of the previous shell, namely the users exclusively retweeting one group and not the other: 
\begin{equation}
    \mathcal{A}_{ANTI}^{(k)}=\mathcal{A}_{ANTI}^{(k-1)}\cup\left(\bigcup_{i\in \mathcal{A}_{ANTI}^{(k-1)}}\mathcal{V}^{IN}_i-\mathcal{A}_{PRO}^{(k)}\right);\qquad  \mathcal{A}_{PRO}^{(k)}=\mathcal{A}_{PRO}^{(k-1)}\cup\left(\bigcup_{i\in \mathcal{A}_{PRO}^{(k-1)}}\mathcal{V}^{IN}_i-\mathcal{A}_{ANTI}^{(k)}\right)
\end{equation}
where $\mathcal{V}^{IN}_i$ represents the set of users retweeting user $i$.\\

The sources of the groups, $\mathcal{S}_{ANTI}^{(k)}$ and $\mathcal{S}_{PRO}^{(k)}$, were identified in the same way using the outgoing links, i.e., users retweeted by the set. The algorithm  converged (quickly here, given the limited diameter of the network) when no new nodes were present in the exclusive neighbourhood. We defined the global audience (and sources) as $k$-shell audiences at the equilibrium.\\
 
Since the number of manually coded activists is too low to perform statistical studies on their behavior (1.4\% of the users in the retweet network), we used the $k$-shell audiences to extend the initial sets. We fixed a threshold on the size of the exclusive neighbourhood being at least 90\% of the new total neighbourhood. We required an overlap between the neighbourhoods such as: 
\begin{equation}
  \#(\mathcal{V}^{IN}_{ANTI}\cap \mathcal{V}^{IN}_{PRO})/\#(\mathcal{V}^{IN}_{ANTI}\cup \mathcal{V}^{IN}_{PRO})<0.l.   
\end{equation}

For this reason, we stopped the procedure at the first iteration ($k=1$) and we defined the new extended activist sets as: $ A_{ext} = \mathcal{A}_{ANTI}^{(1)}$ and $P_{ext}=\mathcal{A}_{PRO}^{(1)}$.
With this procedure we reached  an extended number of classified users covering 23.6\% of the retweet network nodes: 1224 in the $A_{Ext}$ and 2699 in the $P_{Ext}$ set.

\subsection*{{Opinion amplifiers and influencers}}
{In network science, several methods exist to identify the most \emph{central} users, where the meaning of centrality is determined by the research question. We are interested in analyzing two main features: who are the most influential users, namely those who can be considered an authority in the debate; and who contributes the most to amplify information, namely those who participate in producing a flow of information.
For the second part the simplest suitable centrality measure is the out-degree, $k_{out}$, of the nodes: this measure indicates, for a certain user, how many different users she retweeted. \\
Concerning the influencers, the situation is a bit more complex because we must take into account both the number of tweets the user produced and the quantity of retweets she received. We therefore decided to apply an adaptation of the h-index\cite{hirsch2005index} to activity on Twittery: a user has $h-index=k$ if she has $k$ tweets with at least $k$ retweets}. 

{We also idenfied the main hashtags used by both communities. Previous studies\cite{bruns2011use} have emphasized the "central role of the hashtag in coordinating publics"\cite{bruns2015twitter}. Choosing a hashtag both clear enough to be understood, and specific enough to avoid confusions, helps attracting a relevant audience, by increasing one's post "searchability"\cite{zappavigna2015searchable}. Understanding the differentiation between the various actors' tagging strategies is thus key to identify the audience they target, and to assess the potential success they can expect}.

\section*{Acknowledgements}
 This work was supported by a CNRS Momentum grant (to J.K.W.), funds by the Institut Hospitalo-Universitaire (IHU) Méditerranée Infection, the National Research Agency (convention. 15-CE36-0008-01) and the program "Investissements d’avenir" (convention ANR-10-IAHU-03), the Région Provence-Alpes-Côte d'Azur, the European funding FEDER PRIMI, the Institute for Research in Public Health (convention 17-PREV-39) and the ITMO Cancer (Plan cancer 2014-2019). The funding sources had no involvement in the research nor any role in the writing of the manuscript and the decision to submit it for publication. {The authors would like to than the MediaLab team for the data collection}.

\section*{Author contributions statement}
F.G. and J.K.W. designed the experimental setup. with inputs from F.C., P.G.E. and V.S.. F.G. and F.C. performed the analyses.  J.K.W and P.G.E. manually coded the twitter users. All authors participated in the writing of the paper.

\section*{Patient and Public Involvement}
There was no involvement of patients or the public in this research.

\section*{Additional information}
The authors have no competing interests to declare.

The corresponding author is responsible for submitting a \href{http://www.nature.com/srep/policies/index.html#competing}{competing interests statement} on behalf of all authors of the paper. 
The Corresponding Author has the right to grant on behalf of all authors and does grant on behalf of all authors, a worldwide licence to the Publishers and its licensees in perpetuity, in all forms, formats and media (whether known now or created in the future), to i) publish, reproduce, distribute, display and store the Contribution, ii) translate the Contribution into other languages, create adaptations, reprints, include within collections and create summaries, extracts and/or, abstracts of the Contribution and convert or allow conversion into any format including without limitation audio, iii) create any other derivative work(s) based in whole or part on the on the Contribution, iv) to exploit all subsidiary rights to exploit all subsidiary rights that currently exist or as may exist in the future in the Contribution, v) the inclusion of electronic links from the Contribution to third party material where-ever it may be located; and, vi) licence any third party to do any or all of the above. All research articles will be made available on an open access basis. The terms of such open access shall be governed by a Creative Commons licence—details as to which Creative Commons licence will apply to the research article are set out in our worldwide licence referred to above.
Data will be deposited on a public web platform upon publication.

\bibliography{sample}
\end{document}